\documentclass[letterpaper,11pt]{article}
\usepackage[utf8]{inputenc}

\usepackage{amsmath, amsthm, amsfonts, amssymb, thm-restate}
\usepackage{thmtools}
\usepackage{algorithmicx}
\usepackage[table,xcdraw]{xcolor}
\usepackage[ruled,vlined,linesnumbered]{algorithm2e}
\usepackage{setspace}
\usepackage{mathtools}
\usepackage{cite}
\usepackage{comment} 
\usepackage[most]{tcolorbox} 
\usepackage{xfrac}
\usepackage{hyperref}
\usepackage{multirow}
\usepackage{caption}
\usepackage{bm}
\usepackage{newfloat}
\usepackage{enumitem}
\usepackage{dblfloatfix} 
\usepackage{wrapfig}
\usepackage{csquotes}
\usepackage{float}

\usepackage{fancyhdr}
\usepackage{etoolbox} 

\SetCommentSty{mycommfont}
\let\oldnl\nl
\newcommand{\nonl}{\renewcommand{\nl}{\let\nl\oldnl}}

\usepackage[margin=1in]{geometry}
\allowdisplaybreaks

\definecolor{mygreen}{RGB}{10,170,100}
\definecolor{myred}{RGB}{10,170,100}

\hypersetup{
     colorlinks=true,
     citecolor= mygreen,
     linkcolor= myred,
     urlcolor= black
}

\renewcommand{\epsilon}{\varepsilon}

\DeclareMathOperator{\E}{\ensuremath{\normalfont \textbf{E}}}

\newcommand{\hiddencomment}[1]{}

\newcommand{\mc}[1]{\ensuremath{\mathcal{#1}}}

\newcommand{\bone}{\mathbf{1}}

\DeclareMathOperator{\MM}{\mathsf{MM}}

\newcommand{\poly}{\mathrm{poly}}

\usepackage[noabbrev,nameinlink]{cleveref}
\crefname{lemma}{Lemma}{Lemmas}
\crefname{theorem}{Theorem}{Theorems}
\crefname{result}{Result}{Results}
\crefname{property}{Property}{Properties}
\crefname{claim}{Claim}{Claims}
\crefname{definition}{Definition}{Definitions}
\crefname{observation}{Observation}{Observations}
\crefname{proposition}{Proposition}{Propositions}
\crefname{assumption}{Assumption}{Assumptions}
\crefname{line}{Line}{Lines}
\crefname{figure}{Figure}{Figures}
\creflabelformat{property}{(#1)#2#3}
\crefname{equation}{}{}
\crefname{section}{Section}{Sections}
\crefname{appendix}{Appendix}{Appendices}
\crefname{algCounter}{Algorithm}{Algorithms}
\Crefname{algCounter}{Algorithm}{Algorithms}

\newtheorem{theorem}{Theorem}
\newtheorem{result}{Result}
\newtheorem{lemma}{Lemma}[section]
\newtheorem{remark}[lemma]{Remark}

\newcommand{\etal}[0]{\textit{et al.}}

\definecolor{mylightgray}{RGB}{240,240,240}

\algnewcommand{\IIf}[2]{\textbf{if} #1 \textbf{then} #2}
\algnewcommand{\EndIIf}{\unskip\ \algorithmicend\ \algorithmicif}
\algnewcommand{\IElse}[1]{\textbf{else} #1}

\newenvironment{graytbox}{
\par\addvspace{0.1cm}
\begin{tcolorbox}[width=\textwidth,
                  boxsep=5pt,
                  left=1pt,
                  right=1pt,
                  top=2pt,
                  bottom=2pt,
                  boxrule=1pt,
                  arc=0pt,
                  colback=mylightgray,
                  colframe=black,
                  ]
}{
\end{tcolorbox}
}

\newenvironment{whitetbox}{
\par\addvspace{0.1cm}
\begin{tcolorbox}[width=\textwidth,
                  boxsep=5pt,
                  left=1pt,
                  right=1pt,
                  top=2pt,
                  bottom=2pt,
                  boxrule=1pt,
                  arc=0pt,
                  colframe=black,
                  colback=white
                  ]
}{
\end{tcolorbox}
}

\newcounter{algCounter}

\makeatletter
\renewcommand{\paragraph}{%
  \@startsection{paragraph}{4}%
  {\z@}{10pt}{-1em}%
  {\normalfont\normalsize\bfseries}%
}
\makeatother

\makeatletter
\patchcmd{\@algocf@start}
  {-1.5em}
  {0pt}
  {}{}
\makeatother

\newcommand{\Prb}{\mathbb P}
\DeclareMathOperator{\Cov}{Cov}
\DeclareMathOperator{\Var}{Var}
\hypersetup{pdftitle={Polynomial-Degree Sparsification for Stochastic Matching},pdfauthor={}}

\title{\textbf{Stochastic Matching with Polynomial Degree}}
\author{Soheil Behnezhad\thanks{Supported in part by NSF CAREER Award CCF 2442812, a Sloan Fellowship, and a Google Research Award.}\\{\em Northeastern University}}
\date{}

\begin{document}
\maketitle
\thispagestyle{empty}

\thispagestyle{empty}
\begin{abstract}
\setlength{\parskip}{0.2cm}
We study the {\em stochastic matching problem}, where a graph $G$ is known in advance, but each edge survives independently with probability $p$. The goal is to select a subgraph $H$ of small maximum degree before seeing the surviving edges, while preserving a $(1-\epsilon)$ fraction of the expected maximum matching size. This problem has received significant attention due to its numerous applications in kidney exchange, labor markets, and online dating.

In this paper, we show there is always a subgraph $H \subseteq G$ achieving a $(1-\epsilon)$-approximation while having a maximum degree of at most $$O\left(\frac{\log(1/\epsilon)}{\epsilon^4} \cdot \frac{1}{p}\right).$$
This resolves a major conjecture of the literature, first stated by [{\color{mygreen} BDH, STOC'20}], that $\poly(\frac{1}{\epsilon p})$ degrees suffice. The best previous bound was $(1/p)^{\exp((1/\epsilon)\log(1/\epsilon))}$ by [{\color{mygreen} ABGR, STOC'25}].

The key to this result is a lemma of independent interest that any distribution over the matchings of a stochastic graph can be transformed to another distribution which matches each vertex with exactly the same probability, but exhibits extremely low total pairwise correlation.
\end{abstract}

\clearpage
\pagenumbering{arabic}
\section{Introduction}\label{sec:intro}

The stochastic matching problem, introduced by Blum \etal~\cite{Blum15}, is a model for finding large matchings when edge availability is uncertain and must be learned through expensive queries. The problem is motivated by applications in kidney exchange, online labor markets, and online dating. These applications, together with the rich mathematical structure underlying its simple formulation, have led to a flurry of results over the last decade \cite{Blum15,AKL16,AKL17,BR18,AB19,BFHR19,BDFHR19,
GKS19,BDH20,BD20,BBD22,FGH22,ABGR25,DS25}.

Formally, let $G=(V,E)$ be a known graph and let $G_p$
be an (unknown) random subgraph of $G$ retaining every edge independently with probability $p\in(0,1]$.
An algorithm selects a subgraph $H\subseteq G$, queries its edges,
and returns a maximum matching in $H\cap G_p$. Writing $\mu$ for
maximum matching size, the goal is to choose $H$ (without knowing the realization $G_p$) such that
\begin{equation}\label{eq:objective}
 \E\mu(H\cap G_p)\ge(1-\epsilon)\E\mu(G_p).
\end{equation}
While simply taking $H = G$ (i.e., querying all edges) clearly satisfies \cref{eq:objective}, the goal is to keep the maximum degree $\Delta(H)$ of $H$, which corresponds to the number of queries per vertex, small.

\paragraph{Previous results.} The earlier works \cite{Blum15,AKL16,AKL17,BFHR19,AB19} focused on improving the approximation, initially from close to 1/2 and later to 2/3, while keeping the maximum degree dependent only on $p$. The first work to show a $(1-\epsilon)$-approximation is achievable with degree $O_{\epsilon, p}(1)$ was that of Behnezhad, Derakhshan, and Hajiaghayi \cite{BDH20} who achieved so by
$$
\Delta(H)\le
(1/p)^{(\log(1/p))^{\exp(O(1/\epsilon))}}.
$$
They further conjectured that the same algorithm should achieve  $\Delta(H) \leq \poly(1/(\epsilon p))$ which has remained a major open problem since. Partial progress was made by \cite{ABGR25} who obtained
$$
\Delta(H) \leq (1/p)^{\exp((1/\epsilon)\log(1/\epsilon))},
$$
which is $\poly(1/p)$ for fixed $\epsilon$, but far from $\poly(1/(\epsilon p))$.

\paragraph{Main result.} In this paper, we prove that the conjecture of \cite{BDH20} is true.

\begin{graytbox}
\begin{result}[Formally as \cref{thm:short-sampled-union}]
\label{res:main}
For parameters $p \in (0, 1], \epsilon \in (0, 1)$, every graph $G$ has a subgraph $H$ satisfying \cref{eq:objective} while having maximum degree
$$
\Delta(H) \leq O\left( \frac{\log(1/\epsilon)}{\epsilon^4} \cdot \frac{1}{p} \right).
$$
\end{result}
\end{graytbox}
\noindent A few remarks:
\begin{itemize}
    \item To prove \cref{res:main}, we analyze the sample-and-union algorithm of \cite{BFHR19} (formalized as \Cref{alg:sample-and-union}) which is also the algorithm analyzed and conjectured to work by \cite{BDH20}.
    \item In \cref{app:lower-bound} we present a lower bound of $\Omega(\epsilon^{-1} p^{-1})$, which shows that the $p$-dependency of \cref{res:main} is optimal.
    \item The upper bound of \cref{res:main} could be further improved to $O \left( \epsilon^{-2} p^{-1} + \epsilon^{-3} \right)$ using a slightly different algorithm and a more intricate analysis (still based on \cref{lem:covariance}). The main body of this paper presents  the slightly weaker bound of \cref{res:main} for readability and clarity of presentation. We discuss the needed modifications for the improvement in \cref{app:improved-bound}.
\end{itemize}

\section{Technical Overview}
\label{sec:technical-overview}

\paragraph{Crucial and noncrucial edges.}
We begin by recalling the crucial/noncrucial edge decomposition
introduced by Behnezhad, Farhadi, Hajiaghayi, and
Reyhani~\cite{BFHR19}, which has since become a central
ingredient in the analysis of stochastic matching algorithms.
Fix a deterministic maximum matching algorithm $\MM(\cdot)$,
and for every edge $e$, let
\[
q_e:=\Pr[e\in\MM(G_p)]
\]
denote its probability of belonging to the selected maximum
matching of a random realization. For a threshold $\tau>0$,
an edge is called \emph{crucial} if $q_e\ge\tau$, and
\emph{noncrucial} otherwise.

Crucial edges are easy to preserve: since
$\sum_{e\ni v}q_e\le1$, every vertex has at most $1/\tau$
crucial edges, so we can afford to include all of them in
$H$. On the other hand, there may be arbitrarily many
noncrucial edges, but the hope is that they can be used
interchangeably: since each is rarely used by the optimum,
preserving sufficiently many alternatives should suffice.

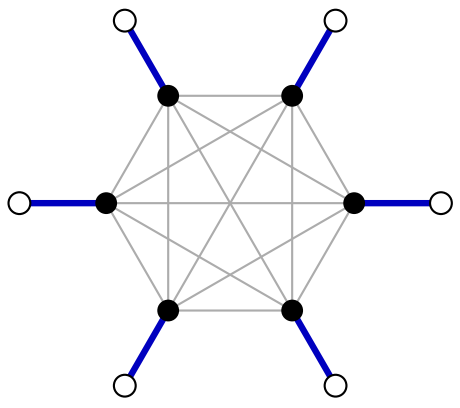
\begin{wrapfigure}{r}{0.19\textwidth}
\vspace{-0.8\baselineskip}
\centering
\begin{tikzpicture}[scale=0.82]
    \foreach \i in {0,...,5} {
        \coordinate (c\i) at ({60*\i}:1.0);
        \coordinate (l\i) at ({60*\i}:1.7);
    }
    \foreach \i in {0,...,5} {
        \foreach \j in {0,...,5} {
            \ifnum\i<\j
                \draw[gray!65] (c\i)--(c\j);
            \fi
        }
    }
    \foreach \i in {0,...,5} {
        \draw[blue!75!black,very thick] (c\i)--(l\i);
        \fill (c\i) circle (2.5pt);
        \draw[fill=white] (l\i) circle (2.5pt);
    }
\end{tikzpicture}
\vspace{-1\baselineskip}
\end{wrapfigure}
The challenge is that both types of edges can contribute
substantially to the optimum, and we must preserve their
contributions simultaneously. A simple example is a large
clique with a private leaf attached to each vertex, as
illustrated on the right. When $p=1/2$, roughly half the
clique vertices can be matched to their leaves, while the
remaining vertices can be matched inside.
Thus, both types of edges contribute substantially,
even though each
individual clique edge has a very small selection probability.

\paragraph{The independence approach.}
A key insight of Behnezhad, Derakhshan, and Hajiaghayi \cite{BDH20} was that we can first commit
to a matching $M_C$ of realized crucial edges, and only
then reveal the realization of the noncrucial edges,
provided that the crucial matching is chosen with
sufficient independence. 

To see why independence matters, consider a noncrucial edge
$e=(u,v)$, and suppose that each of its endpoints is matched
by $M_C$ with probability $1/2$. For us to be able to use
$e$, both endpoints must remain unmatched by $M_C$.
If their matching events were independent, this would
happen with probability $1/4$. However, these events could
be perfectly negatively correlated: for example, exactly
one of $u$ and $v$ might always be matched by $M_C$.
In this case, both vertices are individually unmatched
with probability $1/2$, yet the probability that they
are simultaneously unmatched is zero, meaning that
we can never use the edge $e$.

Unfortunately, full independence is impossible: whenever
a vertex is matched, at least one of its neighbors must
also be matched. The approach of \cite{BDH20} was therefore to construct
a crucial matching in which sufficiently distant vertices
are matched independently. They accomplished this using
a distributed LOCAL algorithm whose output at a vertex
depends only on randomness within a neighborhood of
radius $O_\epsilon(\log\Delta_C)$, where
$\Delta_C$ denotes the maximum degree of the crucial graph.
Consequently, the matching status of a vertex can be
dependent on that of as many as
$
\Delta_C^{O_\epsilon(\log\Delta_C)}
$
other vertices. This quasipolynomial dependence on
$\Delta_C$ is ultimately responsible for the
quasipolynomial dependence on $1/p$ in their result.

The subsequent work of Azarmehr, Behnezhad, Ghafari,
and Rubinfeld~\cite{ABGR25} substantially improved this
approach using local computation algorithms (LCAs).
By carefully bounding both the in-queries and out-queries, they obtained an expected correlated-set
size of only
$
\Delta_C^{\poly(1/\epsilon)}.
$
This improvement was sufficient to obtain a degree
bound polynomial in $1/p$ for every fixed $\epsilon$.
Nevertheless, the exponent of this polynomial still
depends on $\epsilon$.

\paragraph{Our key lemma: bounding total covariance.}
We take a different approach. Rather than requiring
a vertex to be independent of almost every other vertex,
we bound the \emph{total strength of the correlations}
between all pairs of vertices.

Our key lemma shows that every matching distribution can
be transformed into another matching distribution that
preserves the probability of matching each vertex
\emph{exactly}, while having extremely small total
covariance.

\begin{whitetbox}
\textbf{Lemma~\ref{lem:covariance}} (Informal statement). 
Let $M$ be any randomized matching of a stochastic graph
with independent edge realizations. There exists another
randomized matching $M'$ on the same stochastic graph such
that, writing $X_v$ and $X'_v$ for the indicators that
$v$ is matched by $M$ and $M'$, respectively,
$
\E X'_v=\E X_v
$
for every vertex $v$, and
\[
\sum_{u,v}|\Cov(X'_u,X'_v)|=O(\E|M|).
\]
\end{whitetbox}

In particular, the total covariance is bounded by a
constant times the expected number of matched vertices,
with no dependence on the maximum degree of the graph.
This does not mean that each vertex is correlated with
only constantly many others: a vertex may be weakly
correlated with many vertices, but the sum of the
magnitudes of these correlations is small.

\paragraph{An illustrative example.} What makes the lemma particularly surprising is that
it preserves every vertex's matching probability
\emph{exactly}. In fact, there are examples where these
marginals uniquely determine the matching algorithm. To appreciate the strength of this statement, consider a path $P_n$ on vertices $v_1,\ldots,v_n$,
where every edge survives independently with probability
$p$. Let $M$ be the greedy algorithm that processes edges
from left to right, selecting each surviving edge if
the left endpoint is unmatched.

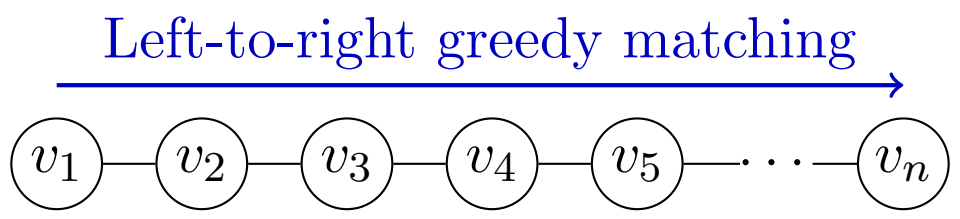
\begin{figure}[h]
\centering
\begin{tikzpicture}[scale=0.8]
    \foreach \i in {1,...,5} {
        \node[circle,draw,minimum size=6mm,inner sep=1pt]
            (v\i) at ({1.2*(\i-1)},0) {$v_{\i}$};
    }
    \node at (5.95,0) {$\cdots$};
    \node[circle,draw,minimum size=6mm,inner sep=1pt]
        (vn) at (7.0,0) {$v_n$};

    \foreach \i in {1,...,4} {
        \pgfmathtruncatemacro{\j}{\i+1}
        \draw (v\i)--(v\j);
    }
    \draw (v5)--(5.65,0);
    \draw (6.25,0)--(vn);

    \draw[->,blue!75!black,thick]
        (0,0.65)--(7,0.65);
    \node[blue!75!black] at (3.5,1.0)
        {Left-to-right greedy matching};
\end{tikzpicture}
\end{figure}

It could be confirmed that the greedy left-to-right algorithm is {\em the only} algorithm that achieves these exact marginals. For example, $(v_1, v_2)$ must be in the matching whenever realized to ensure $v_1$ gets matched with probability $p$. And $(v_2, v_3)$ must be in the matching whenever $(v_2, v_3)$ is realized and $(v_1, v_2)$ is not to make sure the marginal contribution of $v_2$ is satisfied. This means that to preserve the marginals, the proof of \cref{lem:covariance} cannot modify this matching $M$ at all (at least in this example). Instead, we have to show that the total covariance remains remarkable small, despite the apparent propagation of dependencies along this long path.

\paragraph{Proof idea for the key lemma.}
The proof is based on a surprisingly simple optimization
rule: assign each vertex a randomly perturbed weight
\[
w_v=\theta_v+\xi_v,
\]
where $\theta_v$ is a deterministic offset and 
the perturbation $\xi_v \sim \mathsf{Unif}[-1, 1]$ is chosen independently for each vertex.
We then choose a matching maximizing the total weight
of its covered vertices.

There are two main ingredients in the analysis.

\emph{First, we show that the offsets can be chosen to
preserve the prescribed vertex marginals.}
This is far from obvious a priori.
Why should maximizing a linear function of randomly
perturbed vertex weights be capable of reproducing
an arbitrary feasible vector of vertex-matching
probabilities? We prove this using a potential-function argument.
If a vertex is matched less frequently than desired,
we increase its offset; if it is matched too frequently,
we decrease it. The challenge is that changing the weight of a vertex could drastically affect the marginals on other vertices. Nonetheless, we show that these updates make
progress toward the target marginals by decreasing
a suitable potential function. This produces matching distributions whose marginals
can be made arbitrarily close to the desired ones.
Finally, by taking a limit of these distributions,
rather than a limit of the offsets themselves, we
obtain the target marginals exactly.

\emph{Second, we exploit the stability of the
vertex-weighted matching objective.}
The key observation is that changing the weight of
a single vertex, together with any subset of its
incident edges, changes the optimal covered set
at at most two vertices. This stability property is particularly powerful
because all the randomness can be grouped into
independent blocks, one for each vertex.
Resampling any single block then changes the
coverage-indicator vector in at most two coordinates. We analyze the resulting vector of coverage indicators
using a martingale. The small effect of
resampling one block translates into a bound on
the martingale differences, which in turn bounds
the total pairwise covariance.

\section{Key lemma: bounding total covariance}
\label{sec:covariance}

In this section, we prove the key lemma used in the paper.

\begin{whitetbox}
\begin{lemma}\label{lem:covariance}
Let $G(\omega)$ be a random subgraph of a finite graph $G=(V,E)$,
with independent edge indicators; their probabilities need not be
equal. For every randomized matching algorithm $\mc{A}$, there is
another randomized matching algorithm $\mc{A}'$ with the following
property. Write
\[
 M=\mc{A}(G(\omega)),\qquad M'=\mc{A}'(G(\omega)),
 \qquad
 X_v=\bone_{\{v\in V(M)\}},\qquad
 X'_v=\bone_{\{v\in V(M')\}}.
\]
Then $
 \E[X_v]=\E[X'_v]
$
for every vertex $v$,
and
\begin{equation}\label{eq:covariance-lemma}
 \sum_{u,v\in V}|\Cov(X'_u,X'_v)|=O(\E|M|).
\end{equation}
\end{lemma}
\end{whitetbox}

We note that while the lemma preserves vertex marginals, it may not necessarily match edges with the same probability. We write
$s_v:=\E[X_v]$ to the target matching probability of $v$.

\paragraph{Proof outline.} To prove \cref{lem:covariance}, we assign a carefully randomized (and possibly negative) weight to each vertex and let $\mc{B}$ be the matching that maximizes the total weight of covered vertices. In \cref{sec:covariance-bound} prove that maximizing the vertex weights has the desired bound on the total pairwise covariance of all vertices. The harder part of the proof is to show there exists an assignment of weights such that the marginals are satisfied; we prove this in \cref{sec:calibration} using a combinatorial potential-based argument.

\subsection{Stability of vertex weighted matchings}

First, we show that the matching that maximizes vertex weights is stable. This observation and its proof are rather standard in the literature of matching. We present the proof anyway for completeness.

We say a weight vector $w\in\mathbb R^V$ is \emph{generic} if distinct vertex
subsets have distinct weight sums. Note that for generic $w$, every matching $L$
maximizing $\sum_{v\in V(L)}w_v$ has the same covered set. We denote
this set by $S(G,w)$. The optimization is over all matchings, including
the empty matching; it has no additional cardinality constraint.
Weights may be negative.

\begin{lemma}[Stability of the optimal covered set]\label{lem:stability}
Let $w$ be generic. Deleting a vertex $i$ from $G$ either leaves the
optimal covered set unchanged or changes it by the symmetric difference
$\{i,j\}$ for some $j\ne i$. A change is possible only if $i$ is
covered in $S(G,w)$.

More generally, suppose $(G,w)$ and $(G',w')$ differ only in edges
incident to $i$ and in the weight of $i$, and both weight vectors are
generic. Then
\[
 |S(G,w)\mathbin\triangle S(G',w')|\le2.
\]
If the sets differ, at least one of them contains $i$.
\end{lemma}
\begin{proof}
Let $L$ be an optimal matching in $G$ and $L^-$ an optimal matching in
$G-i$. Regard both as matchings on $V(G)$, with $i$ isolated in the
second instance. Their symmetric difference has maximum degree two, and thus decomposes into alternating
paths and even cycles.

Consider a component not containing $i$. Switching the edges of this
component in $L$ is feasible in $G$, and switching it in $L^-$ is
feasible in $G-i$. If the difference between the two covered-weight
sums on this component is $d$, the two optimality inequalities give
$d\ge0$ and $d\le0$. Thus $d=0$. Genericity implies that the two
covered sets on this component agree.

If a component contains $i$, it is a path with endpoint $i$: the
matching $L^-$ has no edge at $i$. Every internal vertex is covered by
both matchings, and only the two endpoints can change coverage. Thus
the covered sets either agree or differ exactly at $i$ and the other
endpoint. If $i$ is uncovered by $L$, there is no such component, and
the sets agree everywhere.

For the second part, delete $i$ from both instances. This leaves
the same graph and the same generic weights. Let $T$ be its unique
optimal covered set. Each original covered set is either $T$ or
$T\mathbin\triangle\{i,j\}$ for some $j\ne i$. Any two sets of these
forms differ at at most two vertices. If neither original set contains
$i$, both equal $T$.
\end{proof}

\subsection{Independent perturbations give the covariance bound}\label{sec:covariance-bound}

Let $\theta\in\mathbb R^V$ be deterministically fixed, and let $\xi_v\sim\mathsf{Unif}[-1,1]$ be chosen independently for each vertex and  also of $\omega$. In $G(\omega)$ choose the (unique) matching $M^\theta$ maximizing the sum of
covered-vertex weights $w_v=\theta_v+\xi_v$. Let $X_v^\theta:=\bone_{\{v\in V(M^\theta)\}}$ be the indicator that vertex $v$ is matched.

We assign each edge of $G$ arbitrarily
to one of its endpoints. Each edge has a realization bit: a Bernoulli
indicator equal to one when that edge is present. Let {\em block} $B_i$ contain
$\xi_i$ and the realization bits of all edges assigned to $i$.
Every bit belongs to exactly one block. Thus the blocks
$B_1,\ldots,B_n$ are independent and together determine $X^\theta$. Note that the coordinates of $X^\theta$ may not be independent.
Resample $B_i$ independently, leaving the other blocks fixed, and write
$X^{\theta,(i)}$ for the resulting vector. Only the weight of $i$ and
some edges incident to $i$ change. By \Cref{lem:stability},
\begin{align}
 \|X^\theta-X^{\theta,(i)}\|_1&\le2,
 \label{eq:resample-uniform}\\
 \E\|X^\theta-X^{\theta,(i)}\|_1
 &\le2\Prb[X_i^\theta=1\text{ or }X_i^{\theta,(i)}=1]
 \le4\E X_i^\theta.
 \label{eq:resample-mean}
\end{align}
The last inequality holds because the two instances have identical distributions.

Let $\mathcal F_i=\sigma(B_1,\ldots,B_i)$ and define the vector-valued
martingale differences
\[
 D_i=\E[X^\theta\mid\mathcal F_i]
       -\E[X^\theta\mid\mathcal F_{i-1}].
\]
Note that $\E[X^{\theta,(i)}\mid\mathcal F_i] = \E[X^\theta\mid\mathcal F_{i-1}]$ since $X^{\theta,(i)}$ resamples the new $B_i$ independently from $\mathcal F_i$. This means
\[
 D_i=\E[X^\theta-X^{\theta,(i)}\mid\mathcal F_i].
\]
Inequalities
\eqref{eq:resample-uniform} and \eqref{eq:resample-mean} give
\begin{equation}\label{eq:martingale-bounds}
 \|D_i\|_1\le2 \qquad \text{and} \qquad
 \E\|D_i\|_1\le4\E X_i^\theta.
\end{equation}
Now from the definition note that 
\begin{equation}
    \sum_{i=1}^n D_i = \sum_{i=1}^n \left( \E[X^\theta\mid\mathcal F_i]
       -\E[X^\theta\mid\mathcal F_{i-1}] \right) = \E[X^\theta\mid\mathcal F_n]
       -\E[X^\theta\mid\mathcal F_{0}] = X^\theta-\E X^\theta.
\end{equation}
From which, we get that
\begin{flalign*}
    \Cov(X_u^\theta,X_v^\theta) = \E[(X_u - \E X_u)(X_v - \E X_v)] = \E \left[ \left( \sum_i D_{i, u} \right) \left( \sum_i D_{i, v}\right) \right] = \sum_{i, j} \E[D_{i, u} D_{j, v}].
\end{flalign*}
The crucial observation here is that for $i \not= j$, we have $\E[D_{i, u} D_{j, v}] = 0$. To see this, suppose w.l.o.g. that $i < j$ and note that
$$
\E[D_{i, u} D_{j, v}] = \E[\E[D_{i, u} D_{j, v} \mid \mathcal F_{j-1}]] = \E[D_{i, u} \E[D_{j, v} \mid \mathcal F_{j-1}]] = \E[D_{i, u} \cdot 0] = 0,
$$
Combining the above, we get that
$
    \Cov(X_u^\theta,X_v^\theta) = \sum_i \E[D_{i,u} D_{i, v}].
$
Consequently,
\begin{align}
 \sum_{u,v}|\Cov(X_u^\theta,X_v^\theta)|
 &=\sum_{u,v}\left|\sum_i\E[D_{i,u}D_{i,v}]\right|\notag\\
 &\leq \sum_{u,v} \sum_i\E[|D_{i,u}| |D_{i,v}|]\notag\\
 &=\sum_i\E\!\left[\left(\sum_v|D_{i,v}|\right)^2\right] \tag{Since $\sum_{u, v} a_u a_v = \left( \sum_u a_u\right) \left( \sum_v a_v \right) = \left( \sum_v a_v \right)^2$.} \\
 &\le2\sum_i\E\|D_i\|_1 \tag{Since $\|D_{i}\|_1 \leq 2$ by \eqref{eq:martingale-bounds}}\\
 &\le8\sum_i\E X_i^\theta.
 \label{eq:fixed-offset-covariance}
\end{align}

\subsection{Satisfying vertex marginals by weight updates}
\label{sec:calibration}

In this section, we show that we could set the deterministic offsets $\theta \in \mathbb{R}^V$ such that the resulting random matching $M^\theta$ of $G(\omega)$ that maximizes the covered vertex weights $w_v = \theta_v + \xi_v$ for independent perturbations $\xi_v \sim \mathsf{Unif}[-1, 1]$ satisfies the marginals of the matching $\mc{A}(G(\omega))$. Namely, let $X$ be the coverage vector of $M=\mc{A}(G(\omega))$ with target marginals $s_v = \E X_v$, and let $X^\theta$ be the coverage vector of $M^\theta$ with vertex marginals $m_v(\theta) := \E X_v^\theta$. We prove in this section that there exists $\theta$ such that $m_v(\theta) = s_v$ for all $v$.

We do so by iteratively updating the offsets. The update has a simple interpretation. If $m_v(\theta)<s_v$, increase
the reward for covering $v$; if $m_v(\theta)>s_v$, decrease it. The challenge is that changing this reward can also change other vertices' coverage probabilities. To analyze this, we define a potential function.

\begin{lemma}[Calibration by single-vertex updates]
\label{lem:combinatorial-calibration}
Let $\rho\in(0,1)$. Start with $\theta=0$.
Whenever $|s_v-m_v(\theta)|>\rho$ for some vertex $v$, make the update
\begin{equation}\label{eq:coordinate-update}
 h:=s_v-m_v(\theta),\qquad
 \theta_v\leftarrow\theta_v+h,
\end{equation}
leaving all other offsets unchanged. Fix a deterministic rule for
choosing a violating vertex. After at most
$\lceil 2n/\rho^2\rceil$ updates, the process stops at a deterministic
vector $\theta^{(\rho)}$ satisfying
\begin{equation}\label{eq:calibration-error}
 \max_v|m_v(\theta^{(\rho)})-s_v|\le\rho,
 \qquad
 \|\theta^{(\rho)}\|_\infty\le\frac{2n}{\rho}.
\end{equation}
\end{lemma}
\begin{proof}
We analyze a change to one coordinate while keeping all other offsets
fixed, and then bound the total number of changes.

\paragraph{The potential.}
For a feasible matching $L$ and fixed perturbations, define its score by
\[
 W_\theta(L):=\sum_{u\in V(L)}(\theta_u+\xi_u).
\]
Define the expected optimal score and the potential by
\begin{equation}\label{eq:calibration-potential}
 F(\theta):=\E W_\theta(M^\theta),\qquad
 \Psi(\theta):=F(\theta)-\sum_u\theta_u s_u.
\end{equation}
On every realized graph, $M$ is a feasible competitor for
$M^\theta$. Since $\xi$ has mean zero and is independent of $X$,
\[
 \E W_\theta(M)=\sum_u\theta_u s_u,
 \qquad
 \Psi(\theta)
 =\E\bigl[W_\theta(M^\theta)-W_\theta(M)\bigr]\ge0.
\]
Also
\begin{equation}\label{eq:initial-potential}
 0\le\Psi(0)=F(0)\le n,
\end{equation}
because at zero offsets every covered vertex contributes at most one.

\paragraph{How one vertex's coverage probability responds.}
Let $\theta'$ agree with $\theta$ except that
$\theta'_v=\theta_v+h$, where $h$ is any real number. Condition on
the realized graph and on all perturbations except $\xi_v$. If some
matching covers $v$, define
\begin{align*}
 A:=\max_{L:\,v\in V(L)}
       \sum_{u\in V(L)\setminus\{v\}}(\theta_u+\xi_u), \qquad\qquad
 B:=\max_{L:\,v\notin V(L)}
       \sum_{u\in V(L)}(\theta_u+\xi_u),
\end{align*}
where both maxima range over matchings in this fixed realized graph.
The empty matching makes the second class nonempty. Every matching
in the first class receives the same contribution $\theta_v+\xi_v$
from $v$; the definition of $A$ removes that contribution. Thus neither
$A$ nor $B$ depends on $\theta_v$ or $\xi_v$, and both remain unchanged
when $\theta_v$ is updated.

The best score in the first class is $\theta_v+\xi_v+A$; the best
score in the second is $B$. Therefore, except on a probability-zero
tie, the optimal matching covers $v$ exactly when
\[
 \xi_v>B-A-\theta_v.
\]
The update changes this threshold by $-h$. Since $\xi_v$ is uniform on
an interval of length two, the conditional coverage probability
changes by at most $|h|/2$. If no matching covers $v$, that probability
is zero under both offsets. Averaging over the conditioning gives
\begin{equation}\label{eq:coordinate-response}
 |m_v(\theta')-m_v(\theta)|\le\frac{|h|}{2}.
\end{equation}

\paragraph{Comparing the old and new optimal scores.}
Use the same realized graph and the same perturbations to choose
$M^\theta$ and $M^{\theta'}$. The matching $M^{\theta'}$ is feasible
under the old weights, so optimality of $M^\theta$ gives
$W_\theta(M^{\theta'})\le W_\theta(M^\theta)$.
The score of the fixed matching $M^{\theta'}$, according to the new weights $\theta'$, only differs from that of $\theta$ by exactly $h$ iff $v$ is matched by $M^{\theta'}$. Therefore,
\begin{equation}\label{eq:fixed-matching-score-change}
 W_{\theta'}(M^{\theta'})-W_\theta(M^{\theta'})
 =h X_v^{\theta'}.
\end{equation}
Here $X_v^{\theta'}$ is one if the particular vertex $v$
is covered by $M^{\theta'}$ and zero otherwise. Thus the score difference is $h$ or
zero: only $v$'s weight changed. Combining the two score
comparisons yields
\begin{align*}
 W_{\theta'}(M^{\theta'})-W_\theta(M^\theta)
 &=\bigl[W_{\theta'}(M^{\theta'})-W_\theta(M^{\theta'})\bigr] +\bigl[W_\theta(M^{\theta'})-W_\theta(M^\theta)\bigr] \le h X_v^{\theta'}.
\end{align*}
Taking expectations, and subtracting the change $h s_v$ in the
target term of the potential, gives
\begin{equation}\label{eq:potential-comparison}
 \Psi(\theta')-\Psi(\theta)
 \le h\bigl(m_v(\theta')-s_v\bigr).
\end{equation}
Note that the inequality holds for either sign of $h$.

\paragraph{Every inaccurate coordinate gives a definite decrease.}
Choose $h=s_v-m_v(\theta)$ as in \eqref{eq:coordinate-update}.
By \eqref{eq:potential-comparison} and \eqref{eq:coordinate-response},
\begin{equation}
 \Psi(\theta')-\Psi(\theta)
 \le h\bigl(m_v(\theta')-m_v(\theta)\bigr)-h^2\notag \le |h|\,|m_v(\theta')-m_v(\theta)|-h^2
 \le-\frac{h^2}{2}.
\end{equation}
Thus every performed update decreases the potential by more than
$\rho^2/2$. The potential starts at most $n$ and is always
nonnegative by \eqref{eq:initial-potential}, so the process must stop within  $\lceil 2n/\rho \rceil$
updates. When the process terminates, no coordinate violates the desired accuracy.
\end{proof}

\paragraph{Getting exact marginals.}
At first, it may seem that applying \Cref{lem:combinatorial-calibration} and letting $\rho_k := 1/k$, $\theta^{(k)} := \theta(\rho_k)$, and taking the limit $\theta^\star = \lim_{k \to \infty} \theta^{(k)}$ would achieve exact marginals. However, note that \Cref{lem:combinatorial-calibration} only gives $\| \theta(\rho) \|_{\infty} \leq 2n/\rho$ which diverges as $\rho \to 0$.

The trick is to take the limit of distributions, and not the weights. We apply
\Cref{lem:combinatorial-calibration} along any sequence
$\rho_k\downarrow0$. For each $k$, use the fixed deterministic offsets
$\theta^{(\rho_k)}$ on a fresh realized graph with independent
perturbations. The random object $(G(\omega), M^{(k)})$ is a joint distribution over a finite space since the graph is finite and the set of matchings is also finite. It therefore has
a convergent subsequence. Its limit has the original graph-realization
distribution, is supported on feasible matchings, and has coverage marginals
exactly $s$ by \eqref{eq:calibration-error}. Define $\mc{A}'$ to
sample from the limiting conditional law of the matching given
$G(\omega)$, using a private independent seed, and write
$M'=\mc{A}'(G(\omega))$ and $X'_v=\bone_{\{v\in V(M')\}}$.
Thus $\E X'_v=s_v=\E X_v$ for every vertex $v$.

\section{Proof of the main theorem using \cref{lem:covariance}}
\label{sec:short-proof}

In this section we prove that the sample-and-union algorithm achieves a $(1-\epsilon)$-approximation with degree $O(\frac{\log(1/\epsilon)}{p\epsilon^4})$. The proof is more or less an adaptation of the arguments of \cite{BDH20,ABGR25} to the total covariance upper bound of \cref{lem:covariance}. 

\begin{theorem}
\label{thm:short-sampled-union}
For every graph $G$, $p\in(0,1]$, and
$\epsilon\in(0,1)$, 
the output $H$ of \Cref{alg:sample-and-union} achieves a $(1-\epsilon)$-approximation with maximum degree $
 R=O\!\left(\frac{\log(1/\epsilon)}{p\epsilon^4}\right).
$
\end{theorem}

\begin{algorithm}[H]
\caption{Sample-and-Union Algorithm}
\label{alg:sample-and-union}
\KwIn{Graph $G=(V,E)$, survival probability $p$, number of samples $R$.}
\KwOut{A subgraph $H\subseteq G$ with maximum degree at most $R$.}

$H\gets (V,\emptyset)$.

\For{$i=1,\ldots,R$}{
    Sample $G^{(i)}\sim G_p$ independently.
    
    Let $M_i = \MM(G^{(i)})$ be a maximum matching of $G^{(i)}$.
    
    $E(H)\gets E(H)\cup M_i$.
}
\Return{$H$}.
\end{algorithm}

\begin{remark}
    \Cref{alg:sample-and-union}, for maximum degree
    $R=O\left(\log(1/\epsilon)(\frac{1}{p\epsilon^2} + \frac{1}{\epsilon^3})\right)$, also could be shown to achieve a $(1-\epsilon)$-approximation using a slightly more complicated analysis. The $\log(1/\epsilon)$ factor could further be removed from this bound by switching to a different algorithm. See \cref{app:improved-bound}.
\end{remark}

\subsection{Notation and elementary tools}
\label{sec:prelim}

\paragraph{Notation.}
All graphs in the main argument are finite, undirected, and simple.
All logarithms are natural. Constants hidden in asymptotic notation
are absolute. 
For a matching $M$, let $V(M)$ be its covered vertices. Write $\mu(K)$
for the maximum cardinality of a matching in $K$ and $\MM(K)$ for the
matching chosen by the fixed rule. A subscript $p$ denotes independent
edge survival with probability $p$. Intersections of graphs refer to
their edge sets on the common vertex set.

For a nonnegative edge vector $z$, use
\[
 |z|:=\sum_e z_e,\qquad
 z_v:=\sum_{e\ni v}z_e,\qquad
 z(F):=\sum_{e\in F}z_e.
\]
The vector is a \emph{fractional matching} if $z_v\le1$ for every
vertex. Let $E_K(S)$ be the edges of $K$ with both endpoints in $S$.
All sums of covariances indexed by $u,v$ are over \emph{ordered} pairs
and include the diagonal.

\paragraph{Tools.} We now mention some elementary tools we use to turn a fractional assignment into an integral 
matching.

\begin{lemma}[A positive-part bound]\label{lem:positivepart}
For $b>0$ and $r\in\mathbb R$,
\[
 (r-b)_+\le\frac{r^2}{4b},\qquad (r)_+:=\max\{r,0\}.
\]
Consequently, if $W$ has finite variance and $c>\E W$, then
\[
 \E(W-c)_+\le\frac{\Var(W)}{4(c-\E W)}.
\]
\end{lemma}
\begin{proof}
The first inequality is immediate for $r\le b$. For $r>b$ it is
equivalent to $0\le(r-2b)^2$. Apply it with
$r=W-\E W$ and $b=c-\E W$, and take expectations.
\end{proof}

\begin{theorem}[Matching polytope; Edmonds~\cite{Edmonds65}]
\label{thm:polytope}
A nonnegative edge vector $z$ on a finite loopless graph is a convex
combination of incidence vectors of matchings if and only if
\[
 z_v\le1\quad\text{for every }v,\qquad
 z(E_K(S))\le\frac{|S|-1}{2}
 \quad\text{for every odd-cardinality }S\subseteq V(K).
\]
In particular, these inequalities imply $\mu(K)\ge|z|$.
\end{theorem}

The following consequence
charges the rounding loss to the sum of squared edge weights.

\begin{lemma}[Rounding with a quadratic loss]\label{lem:quadratic-rounding}
For every fractional matching $y$ in a finite simple graph $K$,
\[
 \mu(K)\ge |y|-\sum_e y_e^2.
\]
\end{lemma}
\begin{proof}
Set $z_e=y_e(1-y_e)$. Since $0\le y_e\le1$, the vector $z$ is
nonnegative and satisfies all vertex constraints.
For an odd set $S$ of size $s\ge3$, put $m=y(E_K(S))\le s/2$.
There are at most $s(s-1)/2$ edges within $S$, so Cauchy--Schwarz gives
\[
 z(E_K(S))
 =m-\sum_{e\in E_K(S)}y_e^2
 \le m-\frac{2m^2}{s(s-1)}.
\]
The expression on the right is nondecreasing on $0\le m\le s/2$.
Consequently,
\[
 z(E_K(S))
 \le\frac s2-\frac{s}{2(s-1)}
 =\frac{s-1}{2}-\frac1{2(s-1)}
 \le\frac{s-1}{2}.
\]
Singleton sets have no internal edges. Thus $z$ satisfies every
constraint of \Cref{thm:polytope}, which gives
$\mu(K)\ge |z|=|y|-\sum_e y_e^2$.
\end{proof}

\begin{lemma}[Repairing vertex loads]\label{lem:local-repair}
For any nonnegative edge vector $x$, put
$y_{uv}=x_{uv}/\max\{1,x_u,x_v\}$. Then $y$ is a fractional matching
on the same support and
\[
 |x|-|y|\le\sum_v(x_v-1)_+.
\]
An edge of weight one whose endpoints both have load one retains
weight one.
\end{lemma}
\begin{proof}
For every vertex $v$, each incident denominator is at least
$\max\{1,x_v\}$, so $y_v\le1$. Assign each reduction $x_e-y_e$ to
an endpoint with larger load. A vertex with load at most one receives
no positive reduction. A vertex with load $x_v>1$ receives at most
$\sum_{e\ni v}x_e(1-1/x_v)=x_v-1$.
Summing these bounds proves the loss inequality. The last assertion
follows directly from the denominator.
\end{proof}

\begin{proof}
Write
\[
 q_e:=\Prb[e\in\MM(G_p)],\qquad
 q_v:=\sum_{e\ni v}q_e\le1,\qquad Q:=\sum_e q_e=\E\mu(G_p).
\]
Choose $0<\delta<1/4$, to be a
sufficiently small absolute constant times $\epsilon$, and set
\begin{equation}\label{eq:short-parameters}
 \gamma:=1-\delta,\qquad \tau:=p\delta^4,\qquad
 R:=\left\lceil\frac{\log(1/\delta)}{\tau}\right\rceil.
\end{equation}
In particular, $e^{-R\tau}\le\delta$ and $1/R\le\tau$.
Let $C:=\{e:q_e\ge\tau\}$ be the set of {\em crucial edges}, $N:=E\setminus C$ the set of {\em noncrucial edges}, and write
$Q_C:=\sum_{e\in C}q_e$,
$q_v^C:=\sum_{e\in C:e\ni v}q_e$, and
$r_v:=\sum_{e\in N:e\ni v}q_e$.

\subsection{The analysis}

\paragraph{The crucial matching $M_C$.} The first step of the proof is an integral matching $M_C$ that we construct on the realized crucial edges $C_p$ using \Cref{lem:covariance}, which guarantees small total pairwise covariance. To do so, we have to specify the target coverage probabilities achievable by a feasible algorithm. Our coverage target is $\gamma q^C_v$. Note that this coverage target is feasible: independently realize noncrucial edges and add them to $C_p$, take the maximum matching (using the same algorithm as in the sample and union algorithm) on the resulting graph and then keep the matching with probability $\gamma$. 

Let $U_v:=\bone_{\{v\notin V(M_C)\}}$ and $a_v:=\E U_v$. We have
\begin{equation}\label{eq:short-availability}
 a_v=1-\gamma q_v^C\ge\delta+\gamma r_v,\qquad
 \E|M_C|=\gamma Q_C,\qquad
 \Gamma:=\sum_{u,w}|\Cov(U_u,U_w)|=O(Q).
\end{equation}
The first inequality holds because $q_v^C = q_v - r_v$ and $q_v \leq 1$. The bound on the covariance holds because letting $X'_v := 1 - U_v$, \Cref{lem:covariance} implies $\sum_{u,w}|\Cov(X'_u,X'_w)|=O(Q)$ and $\Cov(X'_u, X'_v) = \Cov(U_u, U_v)$ since subtracting constants from random variables does not change their covariance. 
Also note that for any vertex $v$,
\begin{align}
\frac{r_v}{a_v} = \frac{q_v-q^C_v}{1 - \gamma q^C_v} \leq q_v,
\end{align}
which implies the following bound that we use 
\begin{equation}\label{eq:short-budget}
 \sum_v\frac{r_v}{a_v}\le\sum_v q_v=2Q.
\end{equation}

\paragraph{A fractional assignment.} Having the (integral) matching $M_C$ on the realized crucial edges, we ``complement'' it by assigning fractional values to the non-crucial edges that are realized and in addition belong to $H$. In particular, for an edge $e$ let
$$
f_e:=R^{-1}\sum_{i=1}^R\bone_{\{e\in M_i\}}
$$
be the fraction of the matchings of \Cref{alg:sample-and-union} that include $e$, which importantly is only supported on $H$. Next, we consider the following assignment of values that is only supported on $H \cap G_p$:
\[
 x_e:=
 \begin{cases}
  \bone_{\{e\in M_C\cap H\}},&e\in C,\\[2pt]
  \displaystyle\gamma\frac{f_e I_e U_uU_v}{p a_u a_v},&e=uv\in N.
 \end{cases}
\]
Here $I_e$ is the indicator that non-crucial edge $e$ is realized, which note is independent of $M_C$ (and thus the pair $(U_u, U_v)$) since $M_C$ is constructed using private randomness on the non-crucial edges.

Every crucial edge is missed by $H$ with probability at most
$e^{-R\tau}\le\delta$, so
$\E|M_C\cap H|\ge\gamma(1-\delta)Q_C$.
For a noncrucial edge $e=uv$, we have
\begin{flalign*}
\E x_{uv}
&=\frac{\gamma}{p a_u a_v}
  \E[f_{uv}I_{uv}U_uU_v]
  \tag{By definition of $x_{uv}$}\\
&=\frac{\gamma}{p a_u a_v}
  \E[f_{uv}]\E[I_{uv}]\E[U_uU_v]
  \tag{By independence}\\
&=\frac{\gamma q_{uv}}{a_u a_v}\E[U_uU_v]
  \tag{Since $\E f_{uv}=q_{uv}$ and $\E I_{uv}=p$}\\
&=\frac{\gamma q_{uv}}{a_u a_v}
  \left(a_ua_v+\Cov(U_u,U_v)\right)
  \tag{By definition of covariance, $\Cov(U_u, U_v) = \E[U_u U_v] - \E[U_u]\E[U_v] = \E[U_u U_v] - a_u a_v$.}\\
&=\gamma q_{uv}
  \left(1+\frac{\Cov(U_u,U_v)}{a_ua_v}\right).
\end{flalign*}
Summing over all noncrucial edges, we get
\begin{align*}
\sum_{uv\in N}\E x_{uv}
&=\gamma Q_N+
\gamma\sum_{uv\in N}
\frac{q_{uv}}{a_ua_v}\Cov(U_u,U_v)\\
&\ge\gamma Q_N-
\gamma\sum_{uv\in N}
\frac{q_{uv}}{a_ua_v}|\Cov(U_u,U_v)|\\
&\ge\gamma Q_N-
\frac{\gamma\tau}{\delta^2}
\sum_{uv\in N}|\Cov(U_u,U_v)|
\tag{Since $q_{uv}<\tau$ and $a_u,a_v\ge\delta$}\\
&\ge\gamma Q_N-\frac{\gamma\tau}{\delta^2}\Gamma\\
&=\gamma Q_N-O\left(\frac{\tau}{\delta^2}Q\right).
\tag{Since $\Gamma=O(Q)$}
\end{align*}
Put together with our earlier bound for crucial edges, we get
\begin{equation}\label{eq:short-mass}
 \E|x|\ge\bigl(1-O(\delta+\tau/\delta^2)\bigr)Q
          =(1-O(\delta))Q.
\end{equation}

\paragraph{From the fractional assignment to a fractional matching.} It has to be noted that the fractional assignment $x$ above is not necessarily a fractional matching because the re-scaling could potentially increase the vertex loads above 1. We now show that this could be avoided.

Let
\[
Y_{vu}:=
\begin{cases}
\displaystyle\frac{f_{vu}I_{vu}U_u}{pa_u},&vu\in N,\\[4pt]
0,&vu\notin N,
\end{cases}
\qquad
Z_v=\sum_u Y_{vu}.
\]
For this calculation, also put $q_{vu}=0$ whenever $vu\notin N$.
The random vectors $f$, $I$, and $U$ are mutually independent,
so $\E Y_{vu}=q_{vu}$ and $\E Z_v=r_v$.
We use the identity
\[
\Var(Z_v)
=\sum_u\Var(Y_{vu})
+\sum_{u\ne w}\Cov(Y_{vu},Y_{vw}),
\]
where the second sum is over ordered pairs.

\paragraph{Individual variances.}
For every $vu\in N$, let
$B_i:=\mathbf{1}_{\{vu\in M_i\}}$.
Note that $B_1,\ldots,B_R$ are independent Bernoulli
random variables with $\E B_i=q_{vu}$, and
$f_{vu}=\frac1R\sum_{i=1}^R B_i$. Therefore,
\begin{align*}
\E f_{vu}^2
&=\frac1{R^2}\E\left[\left(\sum_{i=1}^R B_i\right)^2\right] =\frac1{R^2}\left(
\sum_i\E B_i^2+\sum_{i\ne j}\E[B_iB_j]\right)\\
&=\frac1{R^2}\left(Rq_{vu}+R(R-1)q_{vu}^2\right)
\tag{Since $B_i^2=B_i$ and the $B_i$'s are independent}\\
&=q_{vu}^2+\frac{q_{vu}(1-q_{vu})}{R}\\
&\le\left(\tau+\frac1R\right)q_{vu}.
\tag{Since $q_{vu}<\tau$ and $1-q_{vu}\le1$}
\end{align*}
Now, recalling that
$Y_{vu}=\frac{f_{vu}I_{vu}U_u}{pa_u}$, we have
\begin{align}
\nonumber \Var(Y_{vu})
&=\E Y_{vu}^2-(\E Y_{vu})^2 \le\E Y_{vu}^2 =\frac{\E[f_{vu}^2I_{vu}^2U_u^2]}{p^2a_u^2} =\frac{\E[f_{vu}^2]\E[I_{vu}]\E[U_u]}{p^2a_u^2}
=\frac{\E f_{vu}^2}{pa_u} \\
&\le\frac{\tau+1/R}{p}\frac{q_{vu}}{a_u}.
\end{align}
where we used $\E I_{vu}=p$ and $\E U_u=a_u$ for the last equality and the bound above on $\E f^2_{vu}$ for the last inequality.

\paragraph{Pairwise covariances.}
Consider distinct edges $vu,vw\in N$.
No simulated matching contains both edges. Independence
between different simulations therefore gives
\begin{equation}
\label{eq:Eff}
\E[f_{vu}f_{vw}]
=\frac1{R^2}\sum_{i\ne j}
\Prb[vu\in M_i,\ vw\in M_j]
=\left(1-\frac1R\right)q_{vu}q_{vw}
\le q_{vu}q_{vw}.
\end{equation}
Also, since $vu$ and $vw$ are distinct edges, their test-edge
indicators are independent, and hence
$\E[I_{vu}I_{vw}]=p^2$. By the definition of covariance,
\begin{align*}
\Cov(Y_{vu},Y_{vw})
&=\E[Y_{vu}Y_{vw}]
  -\E[Y_{vu}]\E[Y_{vw}]\\
&=\frac{\E[f_{vu}I_{vu}U_u
                  f_{vw}I_{vw}U_w]}{p^2a_ua_w}
  -q_{vu}q_{vw}
  \tag{By the definition of $Y_{vu}$ and $\E Y_{vu}=q_{vu}$}\\
&=\frac{\E[f_{vu}f_{vw}]
         \E[I_{vu}I_{vw}]
         \E[U_uU_w]}{p^2a_ua_w}
  -q_{vu}q_{vw}
  \tag{Since $f$, $I$, and $U$ are mutually independent}\\
&=\frac{\E[f_{vu}f_{vw}]
         \E[U_uU_w]}{a_ua_w}
  -q_{vu}q_{vw}
  \tag{Since $\E[I_{vu}I_{vw}]=p^2$}\\
&\le\frac{q_{vu}q_{vw}}{a_ua_w}
         \E[U_uU_w]-q_{vu}q_{vw}
  \tag{Since $\E[f_{vu}f_{vw}]\le q_{vu}q_{vw}$ by \eqref{eq:Eff}}\\
&=\frac{q_{vu}q_{vw}}{a_ua_w}
  \left(\E[U_uU_w]-a_ua_w\right)\\
&=\frac{q_{vu}q_{vw}}{a_ua_w}
  \Cov(U_u,U_w)
  \tag{Since $\E U_u=a_u$ and $\E U_w=a_w$}\\
&\le\frac{q_{vu}q_{vw}}{a_ua_w}
  |\Cov(U_u,U_w)|.
\end{align*}
Note that we never assume $U_u$ and $U_w$ are independent.
Their dependence is precisely what the covariance term captures.

\paragraph{Summing over vertices.}
Combining the individual-variance and pairwise-covariance
bounds, rearranging the sums, and adding the nonnegative
terms with $u=w$, we obtain
\[
\sum_v\Var(Z_v)
\le
\frac{\tau+1/R}{p}\sum_u\frac{r_u}{a_u}
+\sum_{u,w}|\Cov(U_u,U_w)|
\sum_v\frac{q_{vu}q_{vw}}{a_ua_w}.
\]
The first sum is at most $2Q$ by \eqref{eq:short-budget}.
For the coefficient of each covariance, we have
\[
\sum_v\frac{q_{vu}q_{vw}}{a_ua_w}
\le\frac{\tau}{a_u}\sum_v\frac{q_{vw}}{a_w}
=\frac{\tau}{a_u}\frac{r_w}{a_w}
\le\frac{\tau}{\gamma\delta},
\]
using $q_{vu}\le\tau$, $a_u\ge\delta$, and
$a_w\ge\gamma r_w$.

Therefore, recalling
$\Gamma=\sum_{u,w}|\Cov(U_u,U_w)|$, we conclude that
\begin{equation}\label{eq:short-variance}
\sum_v\Var(Z_v)
\le\frac{2Q}{Rp}+\frac{2\tau Q}{p}
+\frac{\tau\Gamma}{\gamma\delta}
=O(\delta^3Q).
\end{equation}
The last equality uses $\tau=p\delta^4$, $1/R\le\tau$,
$\Gamma=O(Q)$, and $\gamma\ge3/4$:
the first two terms are $O(\delta^4Q)$, and the third
is $O(p\delta^3Q)=O(\delta^3Q)$.

\paragraph{Controlling the vertex loads.}
Recall that $x_v=\sum_{e\ni v}x_e$ is the total
fractional weight assigned to edges incident to $v$.
We first show that the expected total overload
$\sum_v\E(x_v-1)_+$ is small.

Consider a vertex $v$ covered by $M_C$.
Since $M_C$ is a matching, at most one crucial edge
incident to $v$ receives positive weight, and that weight
is one. Moreover, every noncrucial edge incident to $v$
has weight zero, since $U_v=0$.
Thus $x_v\le1$, so such a vertex has no overload.

Now consider a vertex $v$ not covered by $M_C$.
In this case, $U_v=1$ and no crucial edge incident
to $v$ receives positive weight. Therefore,
\begin{align*}
x_v
&=\sum_{e=vu\in N}x_e =\sum_{e=vu\in N} \gamma\frac{f_eI_eU_u}{pa_ua_v}
 =\frac{\gamma}{a_v}Z_v.
\end{align*}
Consequently, for every vertex $v$, we have
\[
(x_v-1)_+
=U_v\left(\frac{\gamma}{a_v}Z_v-1\right)_+
\le\frac{\gamma}{a_v}
\left(Z_v-\frac{a_v}{\gamma}\right)_+.
\]
Here the inequality follows from $U_v\le1$.

To bound the expectation of the last expression,
recall that $\E Z_v=r_v$ and
$a_v=1-\gamma q_v^C$. Hence,
\begin{align*}
\frac{a_v}{\gamma}-\E Z_v
=\frac{1-\gamma q_v^C}{\gamma}-r_v
=\frac{1-\gamma(q_v^C+r_v)}{\gamma}
=\frac{1-\gamma q_v}{\gamma}
\ge\frac{1-\gamma}{\gamma}
=\frac{\delta}{\gamma}
\end{align*}
where the last inequality uses $q_v\le1$ and $\gamma=1-\delta$.
In particular, the threshold $a_v/\gamma$ is
strictly larger than the mean of $Z_v$.

Recall from \Cref{lem:positivepart} that for any
random variable $W$ and any $c>\E W$,
\[
\E(W-c)_+
\le\frac{\Var(W)}{4(c-\E W)}.
\]
Applying this inequality with $W=Z_v$ and
$c=a_v/\gamma$, we obtain
\begin{align*}
\E(x_v-1)_+
&\le\frac{\gamma}{a_v}
\E\left(Z_v-\frac{a_v}{\gamma}\right)_+\\
&\le\frac{\gamma}{a_v}
\frac{\Var(Z_v)}
{4(a_v/\gamma-\E Z_v)}
\tag{By \Cref{lem:positivepart}}\\
&\le\frac{\gamma^2}{4a_v\delta}\Var(Z_v)
\tag{Since $a_v/\gamma-\E Z_v\ge\delta/\gamma$}\\
&\le\frac1{4\delta^2}\Var(Z_v).
\tag{Since $\gamma\le1$ and $a_v\ge\delta$}
\end{align*}

Summing over all vertices and using
\eqref{eq:short-variance}, we conclude that
\[
\sum_v\E(x_v-1)_+
\le\frac1{4\delta^2}\sum_v\Var(Z_v)
=O(\delta Q).
\]
Thus, although $x$ may violate the fractional
matching constraints $x_v\le1$, the total expected
amount by which these constraints are violated is small.

We now repair these violations by defining
$
y_{uv}:=\frac{x_{uv}}{\max\{1,x_u,x_v\}}.
$
By \Cref{lem:local-repair}, the resulting vector $y$
is a fractional matching, and the total weight lost
during the repair is at most
\[
|x|-|y|\le\sum_v(x_v-1)_+.
\]
Taking expectations and combining this with
\eqref{eq:short-mass}, we obtain
$$
\E|y| \ge\E|x|-\sum_v\E(x_v-1)_+ \ge(1-O(\delta))Q-O(\delta Q) =(1-O(\delta))Q.
$$

Finally, note that every edge in $M_C\cap H$
has weight one in $x$. Its two endpoints have
total load exactly one, since no other crucial
edge can be incident to them and all their
noncrucial edges have weight zero.
Thus the repair does not change these weights:
$y_e=1$ for every $e\in M_C\cap H$.
Moreover, since noncrucial edges have positive
weight only when both endpoints are uncovered
by $M_C$, their support remains vertex-disjoint
from $M_C\cap H$ after the repair.

\paragraph{Rounding.}
We have shown that $y$ is a fractional matching with
$\E|y|\ge(1-O(\delta))Q$. Our final step is to show that
the support of $y$ contains an integral matching of almost
the same size.

Recall from \Cref{lem:quadratic-rounding} that for any
fractional matching $z$ in a simple graph $K$,
\[
\mu(K)\ge |z|-\sum_e z_e^2.
\]
Thus, it suffices to show that the sum of squared weights
of the noncrucial edges is small.

We first bound $\E x_e^2$ for every noncrucial edge $e=uv$.
Recall that we already established
\[
\E f_e^2
=q_e^2+\frac{q_e(1-q_e)}R
\le\left(\tau+\frac1R\right)q_e
\le2\tau q_e,
\]
where the last inequality uses $1/R\le\tau$ from
\eqref{eq:short-parameters}.

By the definition of $x_{uv}$, we have
\begin{align*}
\E x_{uv}^2
&=\E\left[
\left(\gamma\frac{f_{uv}I_{uv}U_uU_v}
{pa_ua_v}\right)^2\right]\\
&=\frac{\gamma^2}{p^2a_u^2a_v^2}
\E[f_{uv}^2I_{uv}U_uU_v]
\tag{Since $I_{uv},U_u,U_v\in\{0,1\}$}\\
&=\frac{\gamma^2}{p^2a_u^2a_v^2}
\E f_{uv}^2\,\E I_{uv}\,\E[U_uU_v]
\tag{By independence of $f$, $I$, and $U$}\\
&=\frac{\gamma^2\E f_{uv}^2}{pa_u^2a_v^2}
\E[U_uU_v]
\tag{Since $\E I_{uv}=p$}\\
&\le\frac{2\tau q_{uv}}{pa_u^2a_v^2}
\E[U_uU_v]
\tag{Since $\E f_{uv}^2\le2\tau q_{uv}$ and $\gamma\le1$}\\
&=\frac{2\tau q_{uv}}{pa_u^2a_v^2}
\left(a_ua_v+\Cov(U_u,U_v)\right)\\
&\le\frac{2\tau q_{uv}}{pa_u^2a_v^2}
\left(a_ua_v+|\Cov(U_u,U_v)|\right).
\end{align*}

Since $y_e\le x_e$ for every edge, we also have
$y_e^2\le x_e^2$. Summing the above inequality
over all noncrucial edges gives
\begin{align*}
\E\sum_{e\in N}y_e^2
&\le\E\sum_{e\in N}x_e^2
\le\frac{2\tau}{p}
\sum_{uv\in N}\frac{q_{uv}}{a_ua_v}
\quad+\frac{2\tau}{p}
\sum_{uv\in N}
\frac{q_{uv}|\Cov(U_u,U_v)|}{a_u^2a_v^2}.
\end{align*}

We bound these two sums separately.
For the first sum, since $a_u,a_v\ge\delta$
by \eqref{eq:short-availability}, we have
\[
\sum_{uv\in N}\frac{q_{uv}}{a_ua_v}
\le\frac1{\delta^2}\sum_{uv\in N}q_{uv}
=\frac{Q_N}{\delta^2}
\le\frac Q{\delta^2}.
\]

For the second sum, we additionally use
$q_{uv}<\tau$ for every noncrucial edge:
\begin{align*}
\sum_{uv\in N}
\frac{q_{uv}|\Cov(U_u,U_v)|}{a_u^2a_v^2}
\le\frac{\tau}{\delta^4}
\sum_{uv\in N}|\Cov(U_u,U_v)|
\le\frac{\tau\Gamma}{\delta^4}.
\end{align*}
The last inequality follows from the definition
$\Gamma=\sum_{u,w}|\Cov(U_u,U_w)|$
in \eqref{eq:short-availability}, since the sum
over noncrucial edges contains only a subset
of the pairs appearing in $\Gamma$.

Combining these two bounds, we obtain
\begin{align*}
\E\sum_{e\in N}y_e^2
&\le\frac{2\tau Q}{p\delta^2}
+\frac{2\tau^2\Gamma}{p\delta^4}
=2\delta^2Q+2p\delta^4\Gamma
=O(\delta^2Q).
\end{align*}
where the first equality follows from $\tau=p\delta^4$ by \eqref{eq:short-parameters} and the second equality follows from  $\Gamma=O(Q)$ by \eqref{eq:short-availability}.
We can now apply \Cref{lem:quadratic-rounding}.
Let $y_N$ denote the restriction of $y$ to
noncrucial edges. Since $y$ is a fractional
matching, $y_N$ is also a fractional matching.
Therefore, the noncrucial support contains
an integral matching $M_N$ satisfying
\[
|M_N|\ge |y_N|-\sum_{e\in N}y_e^2.
\]

Recall that the positive crucial entries of $y$
are exactly the edges of $M_C\cap H$, each with
weight one. Moreover, every noncrucial edge
with positive weight has both endpoints uncovered
by $M_C$. Hence $M_N$ and $M_C\cap H$ are
vertex-disjoint, and their union is a matching
in $H\cap G_p$.

Consequently,
\begin{align*}
\mu(H\cap G_p)
\ge |M_C\cap H|+|M_N|
\ge |M_C\cap H|+|y_N|-\sum_{e\in N}y_e^2
=|y|-\sum_{e\in N}y_e^2.
\end{align*}

Taking expectations and using our bounds
on the repaired fractional matching and
the squared weights, we conclude that
\begin{align*}
\E\mu(H\cap G_p)
\ge\E|y|-\E\sum_{e\in N}y_e^2
\ge(1-O(\delta))Q-O(\delta^2Q)
=(1-O(\delta))Q.
\end{align*}
Recall that $Q=\E\mu(G_p)$.
Choosing $\delta$ to be a sufficiently small
 constant times $\epsilon$ therefore gives
\[
\E_{H,G_p}\mu(H\cap G_p)
\ge(1-\epsilon)\E\mu(G_p).
\]

Finally, by \eqref{eq:short-parameters},
\[
R=\left\lceil\frac{\log(1/\delta)}{p\delta^4}\right\rceil
=O\left(\frac{\log(1/\epsilon)}{p\epsilon^4}\right).
\]
Since $H$ is the union of $R$ matchings, every
output satisfies $\Delta(H)\le R$.
Moreover, averaging over the randomness of $H$
shows that there exists a fixed subgraph $H$
with the same expected approximation guarantee.
This completes the proof.
\end{proof}

\section*{Acknowledgments}

The author thanks Sepehr Assadi, Amir Azarmehr, Avrim Blum, Mahsa Derakhshan, Alireza Farhadi, Alma Ghafari, MohammadTaghi Hajiaghayi, Nima Reyhani, Ronitt Rubinfeld, and Hamed Saleh for many helpful discussions about the stochastic matching problems over the years.

\noindent \textbf{AI Statement:} Discussions with ChatGPT Astra were helpful for this project. In particular, a proof of \cref{lem:covariance} was first completed by ChatGPT after providing relevant context and intuitions. The original proof relied on intricate and hard-to-understand continuous techniques. The author discovered the simpler combinatorial proof that is presented in this paper. The author assumes full responsibility for the correctness of the entire paper and its content.

\appendix

\section{A Lower Bound of $\Omega(1/(p\epsilon))$}
\label{app:lower-bound}

In this section, we show that the dependence on $1/p$ in
\cref{res:main} is optimal. In fact, even on bipartite graphs,
every sparsifier achieving a $(1-\epsilon)$-approximation
must have maximum degree $\Omega(1/(p\epsilon))$.

\begin{theorem}\label{thm:accuracy-lower}
For every $p\in(0,1/2]$ and sufficiently small
$\epsilon>0$, there exists a bipartite graph $G$ such
that every subgraph $H\subseteq G$ satisfying
$\E\mu(H\cap G_p)\ge(1-\epsilon)\E\mu(G_p)$
must have maximum degree
$\Delta(H)=\Omega(1/(p\epsilon))$.
The same lower bound holds for randomized sparsifiers
whose outputs obey the degree bound and are independent
of the test realization.
\end{theorem}

\begin{proof}
Fix $p\le1/2$ and $\epsilon\le1/16$.
Let $k$ be the largest integer satisfying
$(1-p)^k\ge4\epsilon$, and define $a:=(1-p)^k$.
By maximality of $k$ and since $p\le1/2$,
we have $4\epsilon\le a<8\epsilon$.

\paragraph{The construction.}
Consider a complete bipartite graph with $m$ vertices
on each side, which we call the \emph{core vertices}.
Attach $k$ private leaves to each core vertex,
as illustrated below.
Each private leaf is adjacent only to its corresponding
core vertex. Let $G$ denote the resulting bipartite graph.

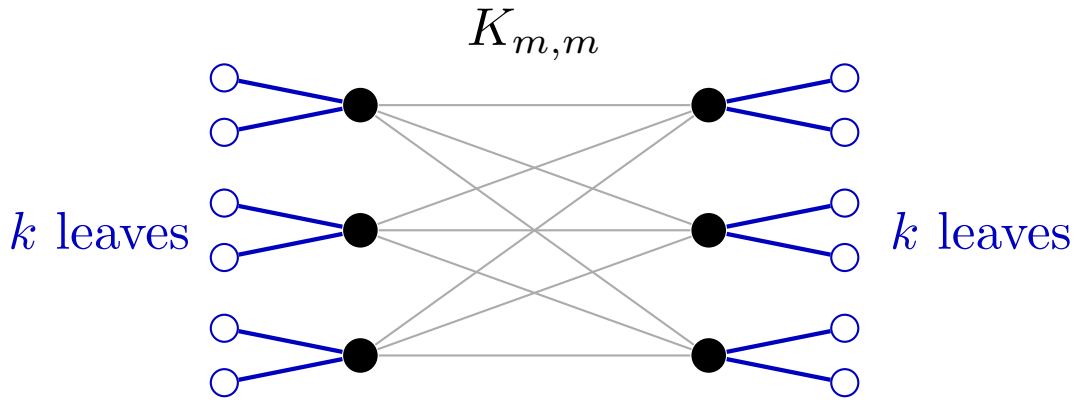
\begin{figure}[H]
\centering
\begin{tikzpicture}[scale=0.72,
    core/.style={circle,fill=black,inner sep=2.3pt},
    leaf/.style={circle,draw=blue!75!black,
                 fill=white,inner sep=1.8pt}
]
    \foreach \i in {0,1,2} {
        \node[core] (L\i) at (0,-1.15*\i) {};
        \node[core] (R\i) at (3.2,-1.15*\i) {};
    }

    \foreach \i in {0,1,2} {
        \foreach \j in {0,1,2} {
            \draw[gray!65] (L\i)--(R\j);
        }
    }

    \foreach \i in {0,1,2} {
        \node[leaf] (LL\i a)
            at (-1.25,{-1.15*\i+0.25}) {};
        \node[leaf] (LL\i b)
            at (-1.25,{-1.15*\i-0.25}) {};
        \draw[blue!75!black,thick] (L\i)--(LL\i a);
        \draw[blue!75!black,thick] (L\i)--(LL\i b);
    }

    \foreach \i in {0,1,2} {
        \node[leaf] (RR\i a)
            at (4.45,{-1.15*\i+0.25}) {};
        \node[leaf] (RR\i b)
            at (4.45,{-1.15*\i-0.25}) {};
        \draw[blue!75!black,thick] (R\i)--(RR\i a);
        \draw[blue!75!black,thick] (R\i)--(RR\i b);
    }

    \node at (1.6,0.65) {\small $K_{m,m}$};
    \node[blue!75!black] at (-2.4,-1.15)
        {\small $k$ leaves};
    \node[blue!75!black] at (5.7,-1.15)
        {\small $k$ leaves};
\end{tikzpicture}
\end{figure}

Call a core vertex \emph{good} if at least one of its
private edges survives, and \emph{bad} otherwise.
Each core vertex is independently bad with probability
$(1-p)^k=a$. Good vertices can be matched to private
leaves, while bad vertices must rely on edges between
core vertices. The full graph contains enough core edges
to match almost all bad vertices, but a bounded-degree
sparsifier may not.

\paragraph{The expected optimum in $G_p$.}
Let $B_L$ and $B_R$ denote the sets of bad core vertices
on the two sides. For every realization, there is a
maximum matching that matches all good vertices to
private leaves and then takes a maximum matching between
the bad vertices. Indeed, replacing an edge incident
to a good vertex by a surviving private edge never
decreases the matching size, and an edge between two
good vertices can be replaced by two private edges.

Consequently, the optimum is
$\mu(G_p)=2m-|B_L|-|B_R|+\mu(G_p[B_L,B_R])$.
Since $|B_L|$ and $|B_R|$ are independent
$\operatorname{Bin}(m,a)$ random variables,
their expected minimum is $ma-O(\sqrt m)$.

Moreover, conditioned on $B_L,B_R$, the edges between
these sets survive independently with probability $p$.
For fixed $p$ and $a$, this random bipartite graph
contains a matching of expected size
$\min\{|B_L|,|B_R|\}-o(m)$.
Indeed, greedily match vertices on the smaller side
while at least $\sqrt m$ unmatched vertices remain
on the other side. Each attempt fails with probability
at most $(1-p)^{\sqrt m}$, so the expected number of
unmatched vertices is at most
$\sqrt m+m(1-p)^{\sqrt m}=o(m)$.

Taking expectations, we conclude that
\begin{equation}\label{eq:lower-full-optimum}
\E\mu(G_p)=m(2-a)-o(m).
\end{equation}

\paragraph{The limitation of bounded-degree sparsifiers.}
Now consider any subgraph $H\subseteq G$ with
maximum degree at most $D$. We give $H$ all private
edges for free, which can only increase its expected
matching size.

Consider a core vertex $v$ with $d_v\le D$ core
neighbors in $H$. The probability that $v$ is bad
is $a$. Conditioned on this event, each of its
$d_v$ core neighbors independently fails to provide
a usable edge with probability $1-pa$: either the
neighbor is good, or the edge does not survive.

Thus, the probability that $v$ is bad and has
no surviving edge to another bad vertex is at least
$a(1-pa)^{d_v}\ge a(1-pa)^D$.
Such a vertex cannot be matched, even after giving
$H$ all private edges for free. Therefore, the expected number of isolated bad core
vertices is at least $2ma(1-pa)^D$.
Since every matching edge between bad vertices
covers two such vertices, we obtain
\begin{equation}\label{eq:lower-sparsifier-optimum}
\E\mu(H\cap G_p)
\le m\left(2-a-a(1-pa)^D\right).
\end{equation}

\paragraph{Concluding the lower bound.}
Suppose $H$ achieves a $(1-\epsilon)$-approximation.
Combining \eqref{eq:lower-full-optimum} and
\eqref{eq:lower-sparsifier-optimum}, and letting
$m\to\infty$, gives
$a(1-pa)^D\le\epsilon(2-a)\le2\epsilon$.
Since $a\ge4\epsilon$, this implies
$(1-pa)^D\le1/2$, and therefore
$D\ge\log2/(-\log(1-pa))$.

Finally, since $pa\le1/4$, we have
$-\log(1-pa)\le2pa\le16p\epsilon$.
Consequently,
\[
D\ge\frac{\log2}{16p\epsilon}
=\Omega\left(\frac1{p\epsilon}\right).
\]

The argument also applies to randomized sparsifiers:
condition on any output $H$, apply
\eqref{eq:lower-sparsifier-optimum}, and then average
over the randomness of $H$.
\end{proof}

\section{Improving the Polynomial Degree Bound}
\label{app:improved-bound}

In this section, we briefly explain how the bound of
\cref{res:main} can be improved to
$
\Delta(H)=O\left(\frac1{p\epsilon^2}
+\frac1{\epsilon^3}\right).
$
The improvement uses the same covariance lemma, but
requires a slightly different sparsification algorithm
and a more refined analysis.

\paragraph{A modified sparsification algorithm.}
Recall that $q_e=\Pr[e\in\MM(G_p)]$.
Set $t=\epsilon/2$, and let
\[
\gamma:=1-t,\qquad
\sigma:=\frac{t^2}{1024},\qquad
\kappa:=\frac{pt^2}{1024},\qquad
\tau:=\min\{\sigma t,\kappa\}.
\]
As before, let $C=\{e:q_e\ge\tau\}$ and
$N=E\setminus C$.

Unlike sample and union, we include \emph{all}
crucial edges directly in $H$. Since
$\sum_{e\ni v}q_e\le1$, this contributes degree
at most $1/\tau$.

For noncrucial edges, we use dependent rounding
\cite{GKPS06}. More precisely, we round the fractional
values $b_e=q_e/\kappa<1$ to indicators $J_e$ with
$\E J_e=b_e$, ensuring that the selected noncrucial
degree at each vertex is at most $1/\kappa+2$.
For general graphs, this can be achieved by orienting
the edges arbitrarily and applying bipartite dependent
rounding to the two-copy graph. The rounding also ensures
nonpositive pairwise covariance between distinct
incident edges within each of the two oriented groups. We return
$
H=C\cup\{e\in N:J_e=1\}.
$ Thus every output satisfies
\[
\Delta(H)\le\frac1\tau+\frac1\kappa+2
=O\left(\frac1{p\epsilon^2}
+\frac1{\epsilon^3}\right).
\]
This construction uses the exact marginals $q_e$;
no efficient exact computation of them is claimed.

\paragraph{Changes to the analysis.}
There are two main improvements over the proof of
\cref{res:main}.

First, we use a stronger consequence of the same
martingale argument behind \cref{lem:covariance}.
The auxiliary crucial matching $M_C$ can be chosen
so that, writing $U_v=\mathbf{1}_{\{v\notin V(M_C)\}}$,
for every family of $L$-Lipschitz functions
$F_v:\{0,1\}^V\to\mathbb R$, we have
\begin{equation}\label{eq:improved-nonlinear}
\sum_v|\Cov(U_v,F_v(U))|
\le16L\,\E|M_C|.
\end{equation}
This follows by the same independent-block exposure
argument: resampling one block changes $U$ in at most
two coordinates, and therefore changes each $F_v$
by at most $2L$. Martingale orthogonality gives the
bound, which is preserved when passing to the limiting
matching distribution with exact marginals.

Second, rather than bounding the overload of every
vertex using the uniform slack $a_v\ge t$, we separate
vertices with $q_v<1/2$ from those with $q_v\ge1/2$.
The former have constant slack, while there are only
$O(Q)$ vertices of the latter type. Together with
\eqref{eq:improved-nonlinear}, this allows us to
control the total repair loss using a total variance
bound of order $O(t^2Q)$, instead of $O(t^3Q)$.

More concretely, define the noncrucial fractional
frequencies by $f_e=\kappa J_e$ and use the same
fractional assignment as in the main proof, with
all crucial edges available. The dependent-rounding
properties give
\[
\E f_e=q_e,\qquad \E f_e^2=\kappa q_e.
\]
Writing $Z_v$ for the corresponding auxiliary loads,
the refined variance calculation gives
\[
\sum_v\Var(Z_v)\le\eta_*Q,
\qquad
\eta_*:=\frac{6\kappa}{p}+16\sigma=O(t^2).
\]

Using \eqref{eq:improved-nonlinear} during repair,
and then applying the same quadratic-rounding lemma,
the resulting relative loss is bounded by
\[
t+\frac{32\sigma}{t}
+2\sqrt{\eta_*}+\frac{\eta_*}{2}
+\frac{\kappa}{pt}
+\frac{16\kappa\sigma}{pt^3}.
\]
With the parameter choices above, this is less than
$\epsilon$.

This proves the improved approximation.
The essential difference is that the stronger
covariance estimate and refined overload analysis
permit a larger noncrucial threshold, while dependent
rounding removes the log factor needed by
sample and union to retain crucial edges.
\qed

\end{document}